\documentstyle[preprint,aps]{revtex}
\tightenlines
\newcommand{\be}{\begin{equation}}
\newcommand{\ee}{\end{equation}}
\newcommand{\ba}{\begin{eqnarray}}
\newcommand{\ea}{\end{eqnarray}}

\begin{document}
\title{Cosmological Constant and the Speed of Light\footnote{This essay received an
``Honorable Mention'' in the Annual Essay Competition of the Gravity Research 
Foundation for the year 2000.}}

\vskip 1.5cm
\author{W. R. Esp\'osito Miguel$^{(1)}$ and J. G. Pereira$^{(2)}$}

\vskip 0.5cm
\address{$^{(1)}$Observat\'orio Municipal de Campinas Jean Nicolini \\
Av. Dr. Heitor Penteado 2145 \\
13087-000\, Campinas SP \\ 
Brazil \\
\vskip 0.5cm
$^{(2)}$Instituto de F\'{\i}sica Te\'orica\\
Universidade Estadual Paulista\\
Rua Pamplona 145\\
01405-900\, S\~ao Paulo SP \\ 
Brazil
}
\maketitle

\begin{abstract}

By exploring the relationship between the propagation of
electromagnetic waves in a gravitational field and the light propagation in a refractive
medium, it is shown that, in the presence of a positive cosmological constant, the velocity
of light will be smaller than its special relativity value. Then, restricting again to the
domain of validity of geometrical optics, the same result is obtained in the context of
wave optics. It is argued that this phenomenon and the anisotropy in the
velocity of light in a gravitational field are produced by the same mechanism. 

\end{abstract}

\newpage

\section{Introduction}

The cosmological constant was introduced by Einstein with the
purpose of obtaining static solutions to the gravitational field equations. Denoting by
$\epsilon \Lambda$ the cosmological constant, with $\Lambda$ positive and $\epsilon = \pm
1$ specifying the sign of the cosmological constant, Einstein's modified equation
reads~\cite{ellis}
\be
R_{\mu \nu} - \frac{1}{2} \, g_{\mu \nu} R +
\epsilon \Lambda g_{\mu \nu} =
\frac{4 \pi G}{c^4} T_{\mu \nu} \; ,
\label{ee1}
\ee
where $T_{\mu \nu}$ is the matter energy-momentum tensor, $G$ is the Newton constant, and
$c$ is the {\it vacuum} speed of light. {\it Vacuum} here means absence of matter
($T_{\mu \nu}=0$) and no cosmological constant ($\Lambda=0$). The basic vacuum solution to
Einstein's equation is the Lorentz metric of (the special relativity) Minkowski spacetime.

In the absence of matter, but with a nonvanishing cosmological constant,
Eq.~(\ref{ee1}) becomes
\be
R_{\mu \nu} - \frac{1}{2} \, g_{\mu \nu} R +
\epsilon \Lambda g_{\mu \nu} = 0 \; ,
\label{ee3}
\ee
whose main solutions are the de Sitter metrics. The introduction of the cosmological
constant, therefore, amounts to replace Minkowski by a de Sitter spacetime as the
spacetime representing absence of matter. They are spacetimes of constant curvature, in
which the Riemann tensor is completely determined by the scalar curvature. Equivalently, as
the scalar curvature $R$ is proportional to $\Lambda$, we can say that in such spacetimes
the Riemann tensor is completely determined by the cosmological constant.

There are two kinds of de Sitter spaces, one with positive, and another one with
negative curvature~\cite{ellis}. They can be defined, respectively, as hypersurfaces
in the pseudo--Euclidean spaces ${\bf E}^{4,1}$ and ${\bf E}^{3,2}$, inclusions whose
points in Cartesian coordinates $\xi^A$ ($A, B, \dots$ = $0, \dots ,4$) satisfy 
\be
\eta_{A B} \, \xi^A \xi^B \equiv \eta_{\mu \nu} \, \xi^{\mu} \xi^{\nu} + \epsilon 
\left(\xi^4\right)^2 = \epsilon {\cal R}^2 \; ,
\label{dspace}
\ee
where ${\cal R}$ is the de Sitter pseudo-radius, $\eta_{\mu \nu}$ ($\mu, \nu, \dots = 
0,1,2,3$) is the Lorentz metric $\eta = $ diag $(-1$, $1$, $1, 1)$, 
and the notation $\eta_{44} = \epsilon$ has been introduced.
 
The de Sitter space $dS(4,1)$, which can be immersed in a pseudo-Euclidean space
${\bf E}^{4,1}$ with metric $\eta_{AB} =$ diag $(-1,+1,+1,+1,+1)$, has the 
pseudo--orthogonal group $SO(4,1)$ as group of motions. It is a 
one-sheeted hyperboloid with topology $R^1 \times S^3$, and positive 
scalar curvature. Therefore, it is a spacetime intrinsically related 
to a positive cosmological constant. The other, which can be immersed in a
pseudo-Euclidean space ${\bf E}^{3,2}$ with metric
$\eta_{AB}$ = diag $(-1,+1,+1,+1,-1)$, is frequently called anti-de Sitter 
space, and is denoted $dS(3,2)$ because its group of motions is 
$SO(3,2)$. The anti-de Sitter space is a two-sheeted hyperboloid with 
topology $S^1 \times R^3$, and negative scalar curvature. It is, 
therefore, a spacetime intrinsically related to a negative 
cosmological constant.

From now on, we are going to consider the case of a positive cosmological constant only.
In stereographic conformal coordinates $x^\mu$~\cite{gursey},
the de Sitter line element
$d\Sigma^2 = \eta_{AB} \, d\xi^A d\xi^B$ is found to be
\be
d\Sigma^2 = g_{\mu \nu} \,d x^\mu dx^\nu \; ,
\label{dsig}
\ee
where
\be
g_{\mu \nu} = \Omega^2(x) \, \eta_{\mu \nu} \; ,
\ee
with $\Omega(x)$ a given function of $x^\mu$. The de Sitter spaces, therefore, are 
conformally flat, with $\Omega^2(x)$ as the conformal factor.

Under an appropriate change of coordinates~\cite{ellis}, the de Sitter line element
$d \Sigma^2$ can be rewritten in the form $(i,j,k,\dots=1,2,3)$\footnote{For the sake of
simplicity, we shall use the same notation $(c t, x^i)$ for the new coordinates.}
\be
d\Sigma^2 = - d \tau^2 + n^2(\tau) \delta_{i j} dx^i dx^j \; ,
\label{dss}
\ee
where
\be
n(\tau) = \exp\left[\textstyle{\sqrt{\frac{\Lambda}{3}}} \, \tau  \right] \; ,
\ee
with $\tau = c \, t$. In these coordinates, therefore, the de Sitter metric is
\be
g_{0 0} = g^{0 0} = - 1 \quad ; \quad
g_{i j} = n^2(\tau) \delta_{i j} \; ,
\label{rime}
\ee
which yields a spacially flat spacetime, with the spacial components of the Ricci tensor
given by
\be
R_{i}{}^{j} = \frac{\Lambda}{3} \; \delta_{i}{}^{j} \; .
\label{ricci2}
\ee

\section{Geometrical Optics Revisited}

In a flat spacetime, the condition for geometrical optics to be applicable is that
\be
\lambda \ll l \; ,
\label{goc}
\ee
where $\lambda$ is the electromagnetic wavelength, and $l$ the typical dimension of
the physical system. Under such condition, any wave-optics quantity $A$ which describes the
wave field is given by a formula of the type
\[
A = a {\rm e}^{i \phi} \; ,
\]
where the amplitude $a$ is a slowly varying function of the coordinates and time, and the
phase $\phi$, called eikonal, is a large quantity which is {\it almost linear} in the
coordinates and the time. The time derivative of $\phi$ yields the angular frequency of the
wave,
\be
\frac{\partial \phi}{\partial t} = - \omega \; ,
\ee
whereas the space derivative gives the wave vector
\be
\frac{\partial \phi}{\partial \mbox{\boldmath$r$}} = \mbox{\boldmath$k$} \; .
\ee

The characteristic equation for Maxwell's equations in an isotropic 
(but not necessarily homogeneous) medium of refractive index $n(r)$ is
\be
\left(\frac{\partial \phi}{\partial \mbox{\boldmath$r$}}\right)^2 -
\frac{n^2(r)}{c^2} \, \left(\frac{\partial \phi}{\partial t}\right)^2 = 0 \; ,
\label{eiko1}
\ee
which implies the usual relation
\be
k^2 = n^2(r) \, \frac{\omega^2}{c^2} \; .
\ee
Without loss of generality, we can assume a characteristic surface in the form
\[
\phi(x, y, z, t) \equiv \frac{\omega}{c} \, \psi(x, y, z) - \omega t = 0 \; ,
\]
and we fall upon the eikonal equation under the form
\be
\left(\frac{\partial \psi}{\partial \mbox{\boldmath$r$}}\right)^2 = n^2(r) \; .
\ee

The wave fronts are surfaces given by $\psi(x,y,z) =$ constant~\cite{luneburg}. A light ray
is defined as a path {\it conjugate} to the wave front in the following sense. If
{\boldmath$r$} is the position vector of a point on the path, with the path length $s$ as
the curve parameter, and
\be
ds = \left(\delta_{i j} \, dx^i dx^j \right)^{1/2}
\label{arc}
\ee
as the length element, then
\be
\mbox{\boldmath$u$} = \frac{d \mbox{\boldmath$r$}}{ds}
\ee
is the tangent velocity normalized to unity. The light ray is then fixed by
\be
n \, \frac{d \mbox{\boldmath$r$}}{ds} =
\frac{\partial \psi}{\partial \mbox{\boldmath$r$}} \; .
\ee
We may eliminate $\psi$ by taking the derivative with respect to $s$. The result
is~\cite{livro}
\be
\frac{d u^k}{d s} - \frac{1}{n} \, \left( \partial^k n -
u^k u^j \partial_j n \right) = 0 \; ,
\label{liray}
\ee
which is the differential equation for the light rays. The equations for 
the light rays are the characteristic equations for the eikonal equations, 
which are themselves the characteristic equations of Maxwell's equations. 
For this reason they are called the {\it bicharacteristic equations} of Maxwell's 
equations. For a homogeneous medium, the refractive index $n$ does not depend on the
space coordinates, and the light-ray equation becomes
\be
\frac{d u^k}{d s} = 0 \; ,
\ee
which represents straight lines.

\section{Light Rays Versus Geodesics}

As is well known, there exists a deep relationship between optical media and
metrics~\cite{gs}. This relationship allows to reduce the problem of the propagation of
electromagnetic waves in a gravitational field to the problem of wave propagation in a
refractive medium in flat spacetime. In fact, for the specific case of the de Sitter
spacetime, with the metric given by Eq.~(\ref{rime}), the curved spacetime eikonal
equation for a $n=1$ refractive medium,
\be
g^{\mu \nu} \, \frac{\partial \phi}{\partial x^\mu} \,
\frac{\partial \phi}{\partial x^\nu} = 0 \; ,
\label{eiko2}
\ee
coincides formally with the flat-spacetime eikonal equation (\ref{eiko1}), valid in a
medium of refractive index $n$. For this reason, $g_{i j}$ is usually called the
{\it refractive metric}.

As a consequence of this relationship, it is possible to show that the light
ray equations (\ref{liray}) for a medium of refractive index $n$ are equivalent to the
3-dimensional geodesics of the refractive metric $g_{i j}$,
\be
\frac{d v^k}{d \tau} + \Gamma^k{}_{i j} \, v^i v^j = 0 \; ,
\ee
with the {\it proper time} $\tau$ defined in Eq.~(\ref{dss}), and $\Gamma^k{}_{i j}$ the
Christoffel of the metric $g_{i j}$. As in the de Sitter spacetime, which is a homogeneous
space, $g_{i j}$ depends only on the time coordinate, $\Gamma^k{}_{i j}$  will be
zero, and the geodesics are simply 
\be
\frac{d v^k}{d \tau} = 0 \; .
\ee
As far as light rays are concerned, the 3-dimensional geodesics correspond to null
geodesics of the de Sitter spacetime. Therefore, we see from (\ref{dss}) and (\ref{arc})
that
\be
d\tau = n(\tau) \, ds \; .
\label{dtds}
\ee

We notice now that there are two velocities at work here. First, there is the
flat-space light-ray velocity
\be
u^j = \frac{d x^j}{d s} \; ,
\ee
which is normalized to unity. Equivalently, due to the
relationship alluded to above, $u^j$ can also be interpreted as the light-ray velocity in
a medium of refractive index $n=1$, that is, as the vacuum light-ray velocity. Second, there
is the light-ray velocity in a de Sitter spacetime,
\be
v^j = \frac{d x^j}{d \tau} \; ,
\ee
which can also be interpreted as the light-ray velocity in a medium of refractive
index $n(\tau)$. These two velocities are related by
$\mbox{\boldmath$v$} = n^{-1}(\tau) \, \mbox{\boldmath$u$}$, or equivalently
$v = n^{-1}(\tau)$. In a dimensional form, therefore, the relation becomes
\be
v = n^{-1}(\tau) \; c \; .
\label{velo}
\ee
As $n(\tau) \geq 1$, we see that, in the presence of a positive cosmological constant,
the velocity of propagation of light will be smaller than its Minkowski (vacuum) value.

\section{Light Waves in a de Sitter Spacetime}

We are going to show now that the same result can be obtained in the context of wave optics,
provided we restrict ourselves to the domain of geometrical optics. Let us then consider
the electromagnetic field equations in a de Sitter spacetime. Denoting the electromagnetic
gauge potential by $A_\mu$, and assuming the generalized Lorentz gauge $\nabla_\mu
A^\mu=0$, with $\nabla_\nu$ the usual Christoffel covariant derivative, the first pair of
Maxwell's equations is
\be
\Box A_\mu - R_{\mu}{}^{\nu} A_\nu = 0 \; ,
\label{me2}
\ee
where $\Box=g^{\lambda \rho} \nabla_\lambda \nabla_\rho$. Since only electromagnetic waves
will be under consideration, we set
\be
A_0 = 0 \; .
\ee
Furthermore, substituting the spacial Ricci tensor components (\ref{ricci2}), Eq. (\ref{me2})
becomes
\be
\Box A_j - \frac{\Lambda}{3} \, A_j = 0 \; .
\label{me3}
\ee
Sometimes, for the specific case of a de Sitter spacetime, the term involving the
cosmological constant is considered as a background-dependent mass term for the
photon field. However, as already discussed in the literature~\cite{faraoni}, this
interpretation leads to properties which are physically unacceptable. In fact, as the
Maxwell equations in four dimensions are conformally invariant, and the de Sitter spaces
are conformally flat, the electromagnetic field must propagate on the
light-cone~\cite{lico}, which implies a vanishing mass for the photon field.

Assuming a massless photon field, therefore, we take the monochromatic plane-wave
solution to the field equation (\ref{me3}) to be
\be
A_j = a_j \exp[i \, K_\mu \, x^\mu] \; ,
\label{mws}
\ee
where $a_j$ is a polarization vector, and
\[
K_\mu = \left(- \frac{\omega(k)}{c}, \mbox{\boldmath$k$} \right)
\]
is the wave-number four-vector, with $\omega(k)$ the angular frequency. In order
to be a solution of (\ref{me3}), the following dispersion relation must be satisfied:
\be
\omega(k) = \frac{c}{n(\tau)} \, \left[ k^2 +
\frac{n^2(\tau) \, \Lambda}{3} \right]^{1/2} \; . 
\label{redi}
\ee
By interpreting
\[
\frac{1}{n(\tau) \, \Lambda^{1/2}} \sim l \; ,
\]
as the typical de Sitter universe dimension, and remembering that $k \sim
\lambda^{-1}$, the condition (\ref{goc}) for geometrical optics to be applicable
turns out to be
\[
k \gg n(\tau) \, \Lambda^{1/2} \; .
\]
In this domain, therefore, the dispersion relation (\ref{redi}) assumes the form
\be
\omega(k) = c \, \frac{k}{n(\tau)}  \; , 
\label{redis}
\ee
and the corresponding velocity of propagation of an electromagnetic wave, given by the
group velocity, is
\be
v \equiv \frac{d \omega(k)}{d k} = n^{-1}(\tau) \; c \; ,
\label{gv}
\ee
which is the same as (\ref{velo}).

\section{Final Remarks}

By considering the domain of geometrical optics, that is, for electromagnetic waves
satisfying the condition
\[
\lambda \ll \frac{1}{n(\tau) \, \Lambda^{1/2}} \; ,
\]
we have shown that, in the presence of a {\it positive} cosmological constant, light
propagates with a velocity smaller than $c$. It is important to remark that it is not $c$
that has a smaller value, but the velocity with which light propagates that is smaller than
$c$, in a way quite similar to the light propagation in a refractive medium. This phenomenon
seems to be produced by the same mechanism that produces the anisotropy in the velocity of
light in a gravitational field~\cite{ohru}. There is a difference, however: As the de Sitter
spacetime is isotropic and homogeneous, the mechanism that produces the anisotropy would,
in this case, act equally in every point and in all directions of
spacetime, yielding an isotropic {\it smaller} velocity of light in relation to the
corresponding special relativity velocity $c$.

There exist several theoretical and experimental indications of a possible non-vanishing
value for the cosmological constant. For example, inflation requires a very high
cosmological constant at the early stages of the universe~\cite{narlikar}. Another example
is the recent measurements~\cite{novae} based 
on observations of very distant type Ia supernovae, which provided a record of the changes
in the expansion rate of the universe over the last several billion years. According to these
observations, a non-vanishing cosmological constant must be at action
to produce the observed rate of expansion. If these indications turn out
to be confirmed, the velocity of light in such Universe might be smaller than its
{\it vacuum} velocity $c$. This result may have important implications to cosmology.

\section*{Acknowledgements}

One of the authors (JGP) would like to thank CNPq-Brazil, for partial
financial support. The other (WREM) would like to thank IFT-UNESP for
the kind hospitality. They would like also to thank R. Aldrovandi for useful
discussions.

\end{document}